\documentclass[dvips,preprint,aps,prb,nofootinbib,endfloats*]{revtex4}
\usepackage[pdftex]{graphicx}
\usepackage{epstopdf}

\newcommand{\itim}{{\sc itim}}
\newcommand{\gitim}{{\sc gitim}}
\usepackage{amsmath,amssymb,amsfonts,mathrsfs}
\begin{document}
\frenchspacing
\newcommand{\preprintclearpage}{\clearpage}

\author{Marcello~Sega}
\email{sega@roma2.infn.it}
\affiliation{Tor Vergata University of Rome, via della Ricerca scientifica 1, I-00133 Rome, Italy}
\author{Sofia~S.~Kantorovich}
\affiliation{Sapienza University of Rome, p.le A. Moro 4, I-00188 Rome, Italy}
\affiliation{Ural Federal University, Lenin Ave. 51, 620083 Ekaterinburg, Russia}
\author{P\'al~Jedlovszky}
\affiliation{Laboratory of Interfaces and Nanosize Systems, Institute of Chemistry, E\"otv\"os Lor\'and University, P\'azm\'any stny 1/a, H-1117 Budapest, Hungary}
\affiliation{MTA-BME Research Group of Technical Analytical Chemistry. Szt. Gell\'ert t\'er 4, H-1111 Budapest, Hungary}
\affiliation{EKF Department of Chemistry, H-3300 Eger, Le\'anyka u. 6, Hungary}
\author{Miguel~Jorge}
\affiliation{LSRE/LCM-Laboratory of Separation and Reaction Engineering, Faculdade de Engenharia, Universidade do Porto, Rua Dr. Roberto Frias, 4200-465 Porto, Portugal}

\title{Identifying interfacial molecules in nonplanar interfaces: the generalized ITIM algorithm}

\begin{abstract}
\noindent{}We present a generalized version of the \itim{} algorithm
for the identification of interfacial molecules, which is able to
treat arbitrarily shaped interfaces. The algorithm exploits the
similarities between the concept of probe sphere used in \itim{}
and the circumsphere criterion used in the $\alpha$-shapes approach,
and can be regarded either as a reference-frame independent version
of the former, or as an extended version of the latter that includes the
atomic excluded volume. The new algorithm is applied to compute
the intrinsic orientational order parameters of water around a DPC
and a cholic acid micelle in aqueous environment, and to the identification of solvent-reachable
sites in four model structures for soot. The additional algorithm
introduced for the calculation of intrinsic density profiles in
arbitrary geometries proved to be extremely useful also for planar
interfaces, as it allows to solve the paradox of smeared intrinsic
profiles far from the interface.
\end{abstract}
\maketitle

\section{Introduction}
Capillary waves represent a conceptual problem for the interpretation
of the properties of liquid-liquid or liquid-vapor planar interfaces,
because long-wave fluctuations are smearing the density profile 
across the interface and all other quantities associated to it. This is usually overcome by
calculating the density profile using a local, instantaneous reference
frame located at the interface, commonly referred to as the intrinsic
density profile, $\rho(z)=\left\langle A^{-1} \sum_{i} \delta\left( z -
z_i\ + \xi(x_i,y_i)\right)  \right\rangle$, where ($x_i$,$y_i$,$z_i$) is the position of the $i$-th atom or molecule, and the local elevation of the surface is 
$\xi(x_i,y_i)$, assuming the
macroscopic surface normal being aligned with the Z axis of a
simulation box with cross section area $A$.  During the last decade
several numerical methods have been proposed to compute the intrinsic
density profiles at
interfaces\cite{chazon03a,chowdhary06a,jorge07a,partay08a,willard10a,jorge10a}. 
Despite several differences in these approaches, they are, in general,
providing consistent distributions of interfacial atoms or molecules\cite{jorge10a} and density profiles\cite{jorge10b}. Among these methods,
\itim{} \cite{partay08a} proved to be an excellent compromise between
computational cost and accuracy\cite{jorge10a}, but it is limited
to macroscopically flat interfaces, therefore there is a need to
generalize it to arbitrary interfacial shapes.

Before these works, albeit for other purposes, several surface-recognition
algorithms have been devised, and will be briefly mentioned below.
All of them are possible starting points for the sought generalization
under the condition that, once applied to the special case of a
planar interface, they lead to consistent results with existing
algorithms for the determination of intrinsic profiles.

Historically, the first class of algorithms addressing the problem
of identifying surfaces was developed to determine molecular areas
and volumes. The study of solvation properties of molecules and
macromolecules (usually, proteins) might require the
identification of molecular pockets, or the calculation of the
solvent-accessible surface area for implicit solvation
models\cite{tomasi05a}. Two intuitive concepts are commonly used
to describe the surface properties of molecules, namely, that of
solvent-accessible surface\cite{lee71a,richards77a}
(SAS), and that of molecular surface\cite{connolly83a,connolly83b}
(MS, also known as solvent excluded surface, or Connolly surface).
The MS can be thought as the surface obtained by letting a hard
sphere roll at close contact with the atoms of the molecule, to
generate a smooth surface made of a connection of pieces of spheres
and tori, which represents the part of the van der Waals surface
exposed to the solvent. During the process of determining the
surface, interfacial atoms can be identified using a simple geometrical
criterion.  Many approximated\cite{shrake73a, richards74a,
greer78a,richmond78a, alden79a, wodak80a, muller83a, pavlov83a,
pascual90a, wang91a, grand93a,weiser99a} or
analytical\cite{connolly83a,connolly83b,connolly85a, richmond84a,
gibson87a,gibson88a,kundrot91a,perrot92a} methods have been
developed to compute the MS or the SAS. In general, these methods
are based on discretization or tessellation procedures, requiring
therefore the determination of the geometrical structure of the
molecule. Other methods which allow to identify molecular surfaces include the approaches of Willard and Chandler\cite{willard10a} or the Circular Variance method of Mezei\cite{mezei03a}. Incidentally, the way the MS is computed in the  early
work of Greer and Bush\cite{greer78a} resembles very closely the
\itim{} algorithm\cite{partay08a}.

From the late 1970s, the problem of shape identification had started
being addressed by a newly born discipline, computational geometry.
In this different framework, several algorithms have been actively
pursued to provide a workable definition of surface, and in particular
the concept of $\alpha$-shapes\cite{edelsbrunner83a,edelsbrunner94a}
showed direct implications for the determination of the molecular
surfaces\cite{edelsbrunner95b,liang98a}. 
The approach based on $\alpha$-shapes is  particularly appealing due
to its generality and ability to describe, besides the geometry,
also the intermolecular topology of the system.

Noticeably, none of these methods -- to the best of our knowledge
-- has ever been employed for the determination of intrinsic properties
at liquid-liquid or liquid-gas interfaces. Prompted by the apparent
similarities between the usage of the circumsphere in the alpha
shapes and that of the probe sphere in the \itim{} method, as we
will describe in the next section, we investigated in more detail
the connection between these two algorithms. As a result, we developed
a generalized version of \itim{} (\gitim{}) based on the $\alpha$-shapes
algorithm. The new \gitim{} method  consistently reproduces the
results of \itim{} in the planar case while retaining the ability
to describe arbitrarily shaped surfaces. In the following we describe
briefly the alpha shapes and the \itim{} algorithms, explain in
detail the generalization of the latter to arbitrarily shaped
surfaces, and present several applications.

\section{Alpha shapes and the generalized \itim{} algorithm}
\begin{figure}[t!]
\raisebox{-0.5\height}{\includegraphics[width=0.68\columnwidth]{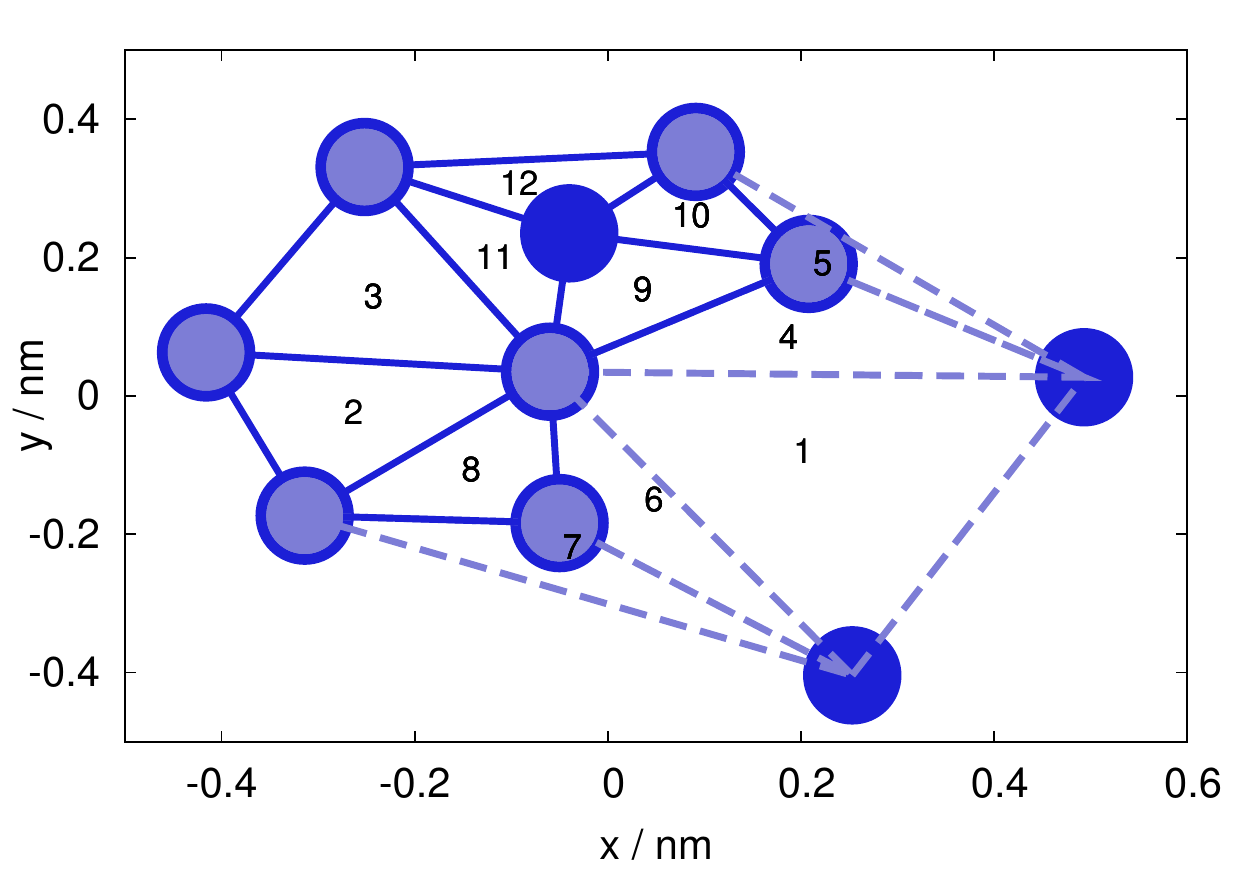}}
\raisebox{-0.45\height}{\includegraphics[width=0.30\columnwidth]{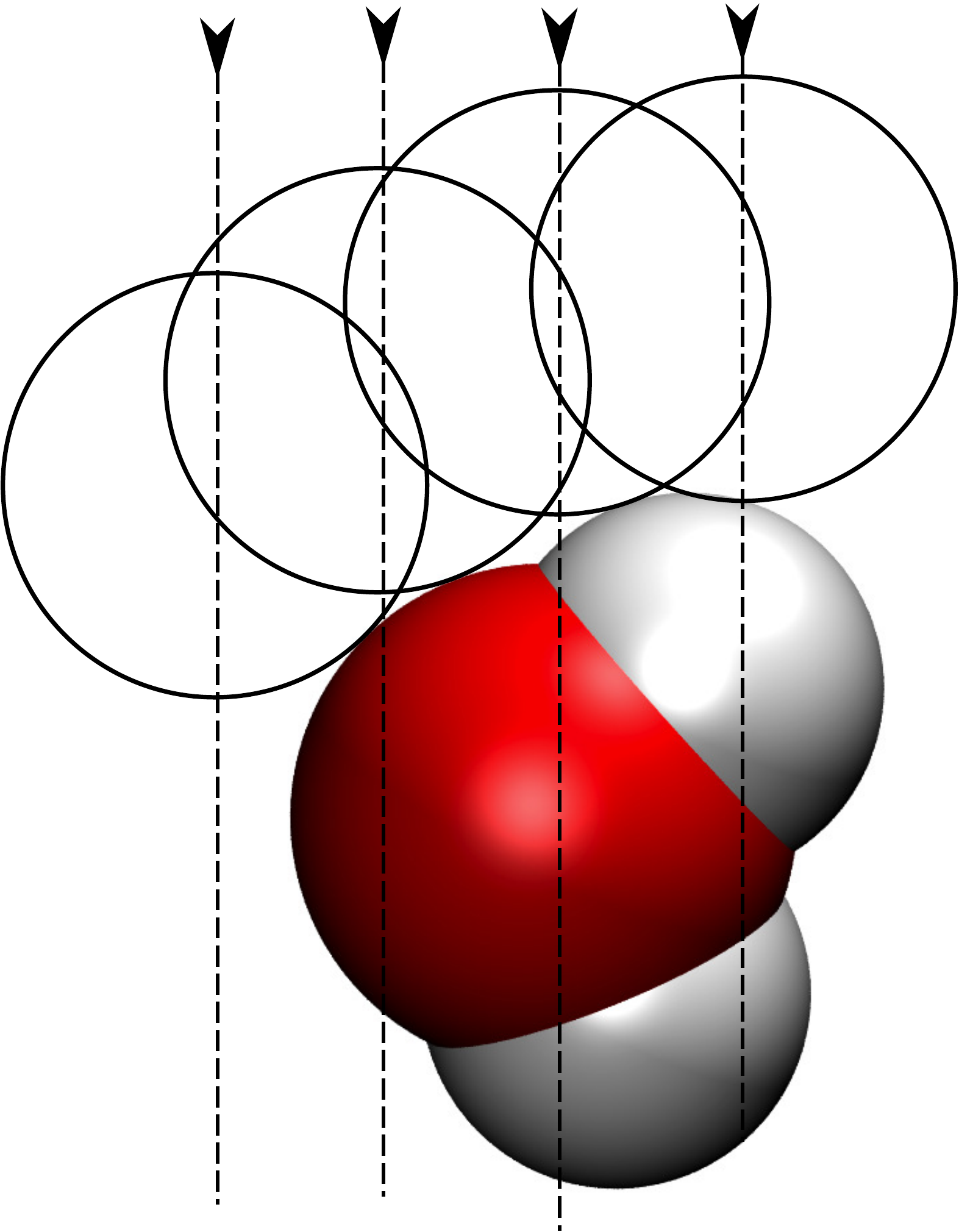}}
\caption{Left: example of the $\alpha$-shapes algorithm on a set of points on the plane. The lines connecting the atoms represent the Delaunay triangulation (the triangles are labeled by numbers from 1 to 12). Solid lines mark triangles belonging to the $\alpha$-complex, and dashed lines those which are not. The light-shaded atoms are those belonging to the $\alpha$-shape, the border of the $\alpha$-complex. Two atoms (in triangle 1) are outside the $\alpha$-shape, and one (shared by triangles 9-12) is inside the $\alpha$-shape. Right: schematic representation of the \itim{} algorithm, applied to a single water molecule: the probe spheres (circles) are moved down the test lines (dashed lines) until they touch an atom.
\label{fig:sketch}}
\end{figure}

The concept of $\alpha$-shapes was introduced several decades ago by
Edelsbrunner\cite{edelsbrunner83a,edelsbrunner94a}. To date the
method is applied in computer graphics application for digital shape
sampling and processing, in pattern recognition algorithms and in
structural molecular biology\cite{edelsbrunner04a}.
The starting point in the determination of the surface of a set of
points in the $\alpha$-shapes algorithm is the calculation of the
Delaunay triangulation, one of the most fruitful concepts for
computational geometry\cite{delaunay34a,okabe00a}, which can be defined in
several equivalent ways, for example, as the triangulation that
maximizes the smallest angle of all triangles, or the triangulation 
of the centers of neighboring Voronoi cells. The idea behind the
$\alpha$-shapes algorithm is to perform a Delaunay triangulation of a
set of points, and then generate the so-called $\alpha$-complex from
the union of all k-simplices (segments, triangles and tetrahedra,
for the simplex dimension k=1,2 and 3, respectively), characterized by a k-circumsphere
radius (which is the length of the segment, the radius of the
circumcircle and the radius of the circumsphere for k=1,2 and 3,
respectively) smaller than a given value, $\alpha$ (hence the name).
The $\alpha$-shape is then defined as the border of the $\alpha$-complex,
and is a polytope which can be, in general, concave, topologically
disconnected, and composed of patches of triangles, strings of edges
and even sets of isolated points.
In a pictorial way, one can imagine the $\alpha$-shape procedure
as growing probe spheres at every point in space until they
touch the nearest four atoms. These spheres will have, in general,
different radii. Those atoms that are touched by spheres with radii
larger than the predefined value $\alpha$ are considered to be at
the surface.

\begin{figure}[t]
\includegraphics[width=\columnwidth]{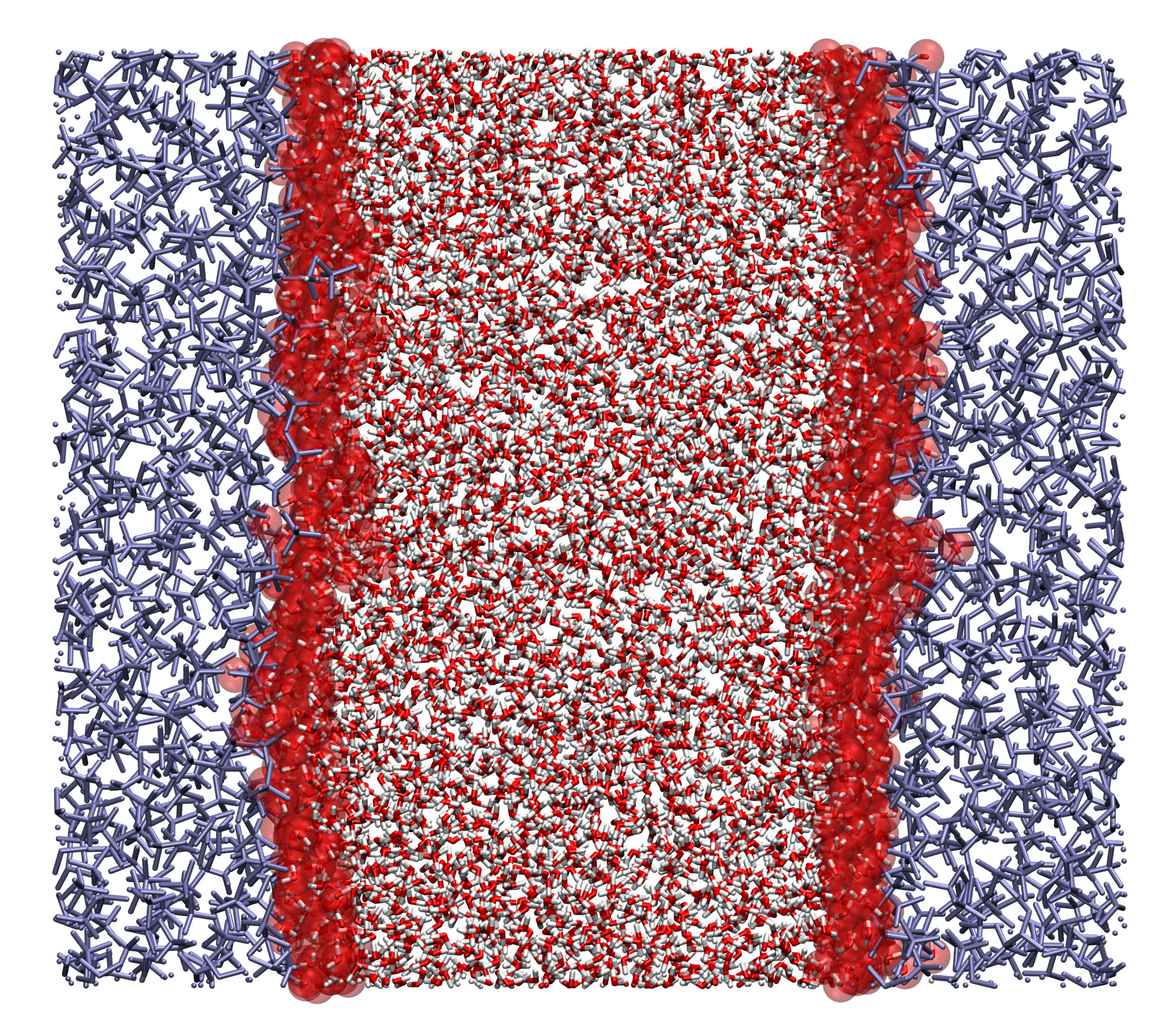}
\caption{Simulation snapshot of a H$_2$O/CCl$_4$ system. The oxygen atoms at the interface between the H$_2$O phase (inner) and CCl$_4$ phase (outer) as recognized by the \gitim{} algorithm are represented with an additional halo. Unconnected points belong to molecules which cross periodic boundary conditions.\label{fig:mixture}}
\end{figure}

An example of the result of the $\alpha$-shapes algorithm in two dimensions
is sketched in Fig.~\ref{fig:sketch}a. The \itim{} algorithm is based instead on the
idea of selecting those atoms of one phase that can be reached by
a probe sphere with fixed radius streaming from the other phase
along a straight line, perpendicular to the macroscopic surface.
An atom is considered to be reached by the probe sphere if the two
can come at a distance equal to the sum of the probe sphere and
Lennard-Jones radii, and no other atom was touched before along the
trajectory of the probe sphere. In practice, one selects a finite
number of streamlines, and if the space between them is considerably smaller
than the typical Lennard-Jones radius $R_p$, the result of the algorithm
is practically independent of the location and density of the streamlines.  The
same is not true regarding the orientation of the streamlines; this
is a direct consequence of the algorithm being designed for
planar surfaces only. The basic idea behind the \itim{}
algorithm are sketched in Fig.~\ref{fig:sketch}b. A closer
inspection reveals that the condition of being a surface atom for
the \itim{} algorithm resembles very much that of the $\alpha$-shapes
case. Quadruplets of surface atoms identified by the \itim{} algorithm
have the characteristic of sharing a common touching sphere having
the same radius as the probe sphere. In this way, one can see the
analogy with the $\alpha$-shapes algorithm, the $R_p$ parameter
being used instead of $\alpha$. The most important differences in
the $\alpha$-shapes algorithm with respect to \itim{} are the absence
of a volume associated with the atoms, and its independence from any
reference frame. We devised, therefore, a variant of the $\alpha$-shapes
algorithm that takes into account the excluded volume of the
atoms.

In the approach presented here the usual Delaunay
triangulation is performed, but the $\alpha$-complex is computed
substituting the concept of the circumsphere radius with that of
the radius of the touching sphere, thus introducing the excluded
volume in the calculation of the $\alpha$-complex. Note that this is
different from other approaches that are trying to mimic the
presence of excluded volume at a more fundamental level, like the
weighted $\alpha$-shapes algorithm, which uses the so-called regular
triangulation instead of the Delaunay one\cite{edelsbrunner94a}.
In addition, in order to eliminate all those complexes, such as
strings of segments or isolated points, which are rightful elements
of the shape, but do not allow a satisfactory definition of a
surface, the search for elements of the $\alpha$-complex stops in
our algorithm at the level of tetrahedra, and triangles and segments
are not checked. In this sense \gitim{} can provide substantially
different results from the original $\alpha$-shapes algorithm.

The equivalent of the $\alpha$-complex is then realized by selecting the 
tetrahedra from the Delaunay triangulation whose touching sphere is smaller than
a probe sphere of radius $R_p$, and the equivalent of the $\alpha$-shape
is just its border, as in the original $\alpha$-shapes algorithm.
The procedure to compute the touching sphere radius is described
in the Appendix.

In the implementation presented here, in order to compute efficiently
the Delaunay triangulation, we have made use of the quickhull
algorithm, which takes advantage of the fact that a Delaunay
triangulation in $d$ dimensions can be obtained from the ridges of
the lower convex hull in $d+1$ dimensions of the same set of points
lifted to a paraboloid in the ancillary dimension\cite{brown79a}.
The quickhull algorithm employed here\cite{barber96a} has
the particularly advantageous scaling $\mathcal{O}(N\log(\nu))$ of
its computing time with the number $N$ and $\nu$ of input points
and output vertices, respectively.  

A separate issue is represented by the calculation of the intrinsic
profiles (whether profiles of mass density or of any other quantity)
as the distance of an atom in the phase of interest from the surface
is not calculated as straightforwardly as in the respective
non-intrinsic versions. For each atom in the phase, in fact, three
atoms among the interfacial ones have to be identified in order to
determine by triangulation\cite{jorge10b} the instantaneous, local position of the
interface. This issue will be discussed in Sec.~\ref{sec:comparison}
for the planar, for the spherical or quasi-spherical and for the
general case: here we simply note that we turned down an early
implementation of the algorithm that searches for these surface atoms,
based on the sorting of the distances using $\mathcal{O}(N\log
N)$  algorithms like quicksort, in favor of a better performing 
approach, based on kd-trees\cite{bentley75a,bentley79a}, a generalization of
the one-dimensional binary tree,  which are still built in a
$\mathcal{O}(N\log N)$ time, but allow for range search in (typically)
$\mathcal{O}(\log N)$ time.

\section{Comparison between the \itim{} and  the \gitim{} methods\label{sec:comparison}}
We have compared the results of the \itim{} and \gitim{} algorithms
applied to the water/carbon tetrachloride interface
composed of 6626 water and 966  CCl$_4$ molecules. The
water and CCl$_4$ molecules have been described by the TIP4P model\cite{jorgensen83a},
and by the potential of McDonald and
coworkers\cite{mcdonald82a}, respectively. The molecules have been kept rigid using the
SHAKE algorithm\cite{ryckaert77a}. This simulation, as well as the
others reported in this work have been performed using the
Gromacs\cite{hess08a} simulation package employing an integration
time step of 1 fs, periodic boundary conditions, a cutoff at 0.8 nm
for Lennard-Jones interactions and  the smooth Particle Mesh Ewald
algorithm\cite{essmann95a} for computing the electrostatic interaction,
with a mesh spacing of 0.12 nm (also with a cut-off at 0.8 nm for the real-space part of the interaction).  All simulations were performed in
the canonical ensemble at a temperature of 300K using the Nos\'e--Hoover
thermostat\cite{nose84a,hoover85a} with a relaxation time of 0.1
ps. A simulation snapshot of the H$_2$O/CCl$_4$ interface is presented
in Fig.~\ref{fig:mixture}, where the surface atoms identified by the
\gitim{} algorithm using a probe sphere radius of 0.25 nm are
highlighted using a spherical halo.

\begin{figure}[t!]
\includegraphics[width=\columnwidth]{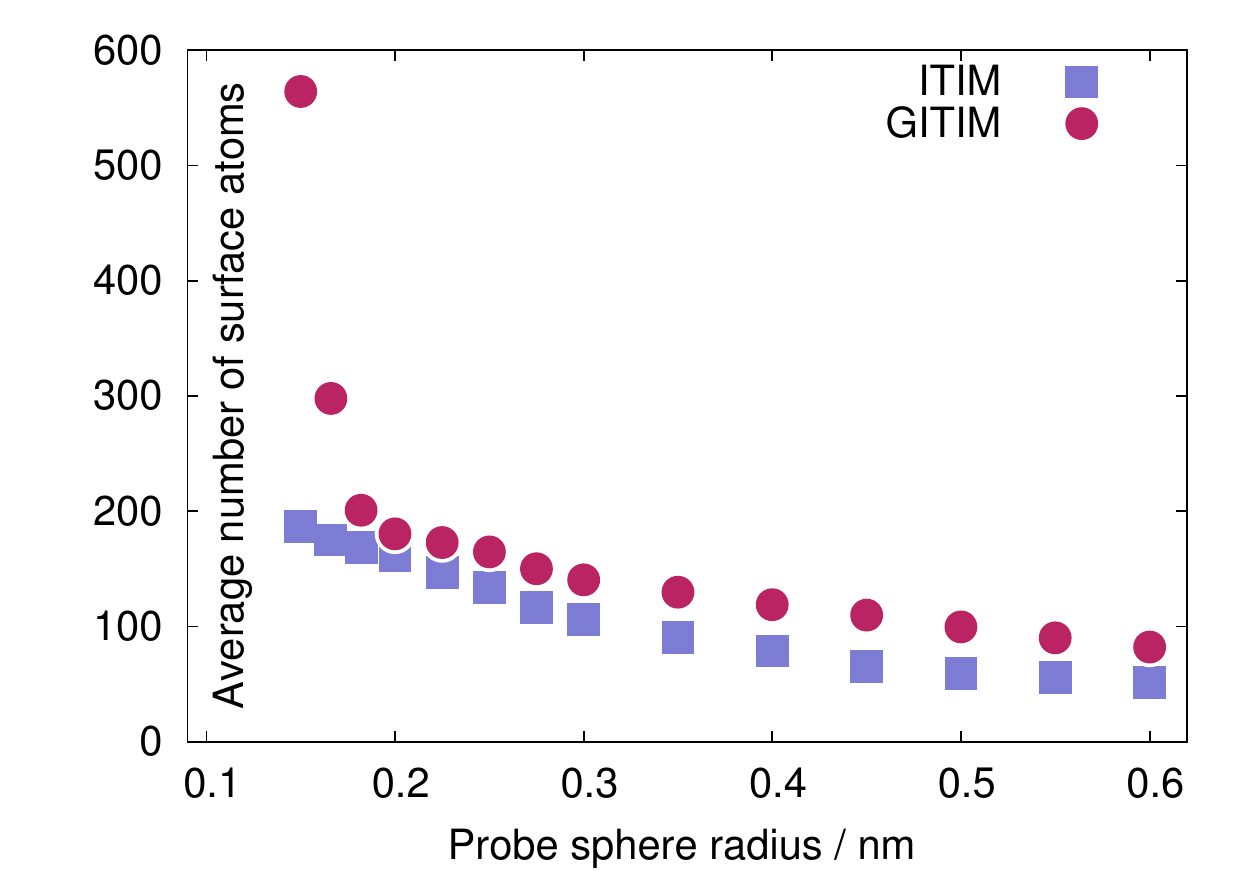}
\caption{Average number of surface atoms identified by \itim{} (squares) and \gitim{} (circles) as a function of the probe sphere radius.\label{fig:NvsA}}
\end{figure}

We have used the \itim{} and \gitim{} algorithms to identify the interfacial
atoms of the water phase in the system, for different sizes of the
probe sphere. In general, \gitim{} identifies systematically a
larger number of interfacial atoms than \itim{}  for the same value
of the probe sphere radius $R_p$, as it is clearly seen in
Fig.~\ref{fig:NvsA}.  Remarkably, for values of the probe sphere radius smaller than about 0.2 nm
(compare, for example, with the optimal \itim{} parameter $R_p=0.125$ nm suggested in
Ref.~\onlinecite{jorge10a}), the interfacial atoms identified by
\gitim{} show the onset of percolation.  The reason for this behavior traces back to the
fact that \itim{} is unable to identify voids buried in the middle
of the phase, as it is effectively probing only the cross section
of the voids along the direction of the streamlines. 
This difference could explain the higher number of surface atoms
identified by \gitim{}, as voids in a region with high local curvature
(or, in other words, with a local surface normal which deviates
significantly from the macroscopic one) will not be identified as such
by \itim.  In \gitim, on the contrary, probe spheres can be thought
as inflating at every point in space instead of moving down
the streamlines, and this is the reason why the algorithm is able
to identify also small pockets inside the opposite phase.

It is possible to make a rough but enlightening analytical estimate of the probability for a
probe sphere of null radius in the \itim{} algorithm to penetrate for a distance $\zeta$ in
a fluid of hard spheres with diameter $\sigma$ and number density
$\rho$. Using the very crude approximation of randomly distributed
spheres, the probability $p_0$ to pass the first molecular layer,
at a depth $\zeta=\sigma$ is the effective cross section
$p_0=1-\frac{\pi}{4}\rho^{2/3} \sigma^2 $, and that of reaching a
generic depth $\zeta$ can be approximated as $p(\zeta) =
p_0^{\zeta/\sigma}$, where $\kappa=\ln(1/p_0)/\sigma$ defines a
penetration depth.  Therefore, using a probe sphere with a  null
radius, \itim{} will identify a (diffuse) surface at a depth
$1/\kappa$, while \gitim{} will identify every atom as a surface
one. For water at ambient conditions, the  penetration
is $\kappa^{-1}\simeq 0.186$ nm, a distance smaller than the size
of a water molecule itself. This could explain why in Ref.~\onlinecite{jorge10a},
even using a probe sphere radius as small as 0.05 nm, almost only
water molecules in the first layer were identified as interfacial
ones by \itim{} (see the almost perfectly Gaussian distribution of
interfacial water molecules in Fig.9 of Ref.~\onlinecite{jorge10a}).

\begin{figure}[t!]
\includegraphics[width=\columnwidth]{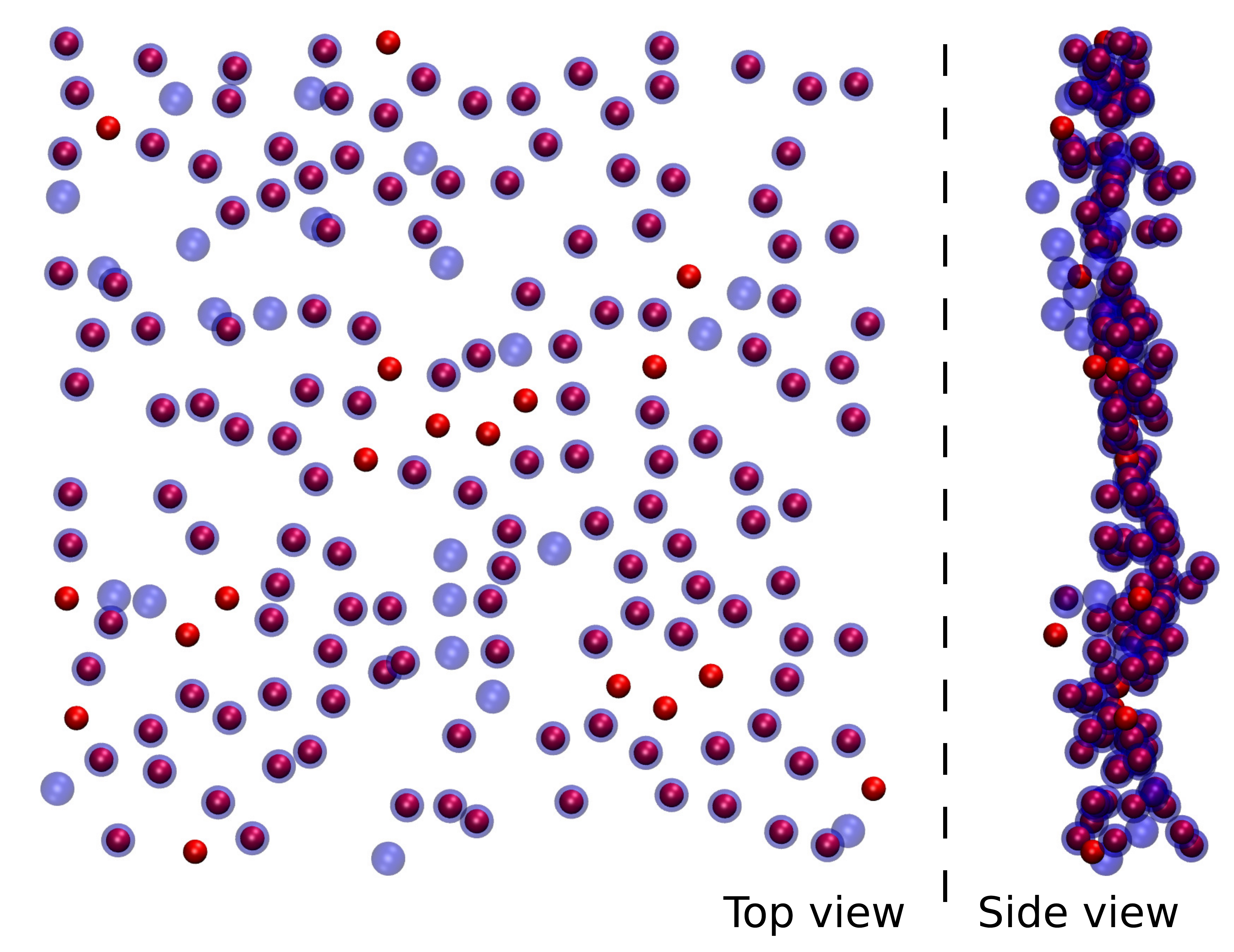}
\caption{Water surface oxygen atoms in the H$_2$O/CCl$_4$ system in one simulation snapshot as recognized by \gitim{} exclusively (small spheres), \itim{} exclusively (large spheres) or by both methods (sphere with halo).\label{fig:recognition}}
\end{figure}

Nevertheless, it is important for practical reasons to be able to
match the outcome of both algorithms. It turns out that choosing
$R_p$ so that the average number of interfacial atoms identified
by both algorithms is roughly the same leads also, not surprisingly,
to very similar distributions. The probe sphere radius required for
\gitim{} to obtain a similar average number of surface atom as in
\itim{} can be obtained by an interpolation of the values reported
in Fig.~\ref{fig:NvsA}. An example showing explicitly the interfacial atoms
identified by the two methods ($R_p=0.2$ nm for \itim{} and $R_p=0.25$
nm for \gitim{}) is presented in Fig.~\ref{fig:recognition}: roughly
85\% of surface atoms are identified simultaneously by both methods,
demonstrating the good agreement between the two methods once the
probe sphere radius has been re-gauged. 
The condition of identifying the same atoms as interfacial ones is
much more strict that any condition on average quantities, like the
spatial distribution of interfacial atoms or intrinsic density
profiles. Hence, it is expected that a good agreement on such
quantities can also be achieved.

\begin{figure}[t]
\includegraphics[width=\columnwidth]{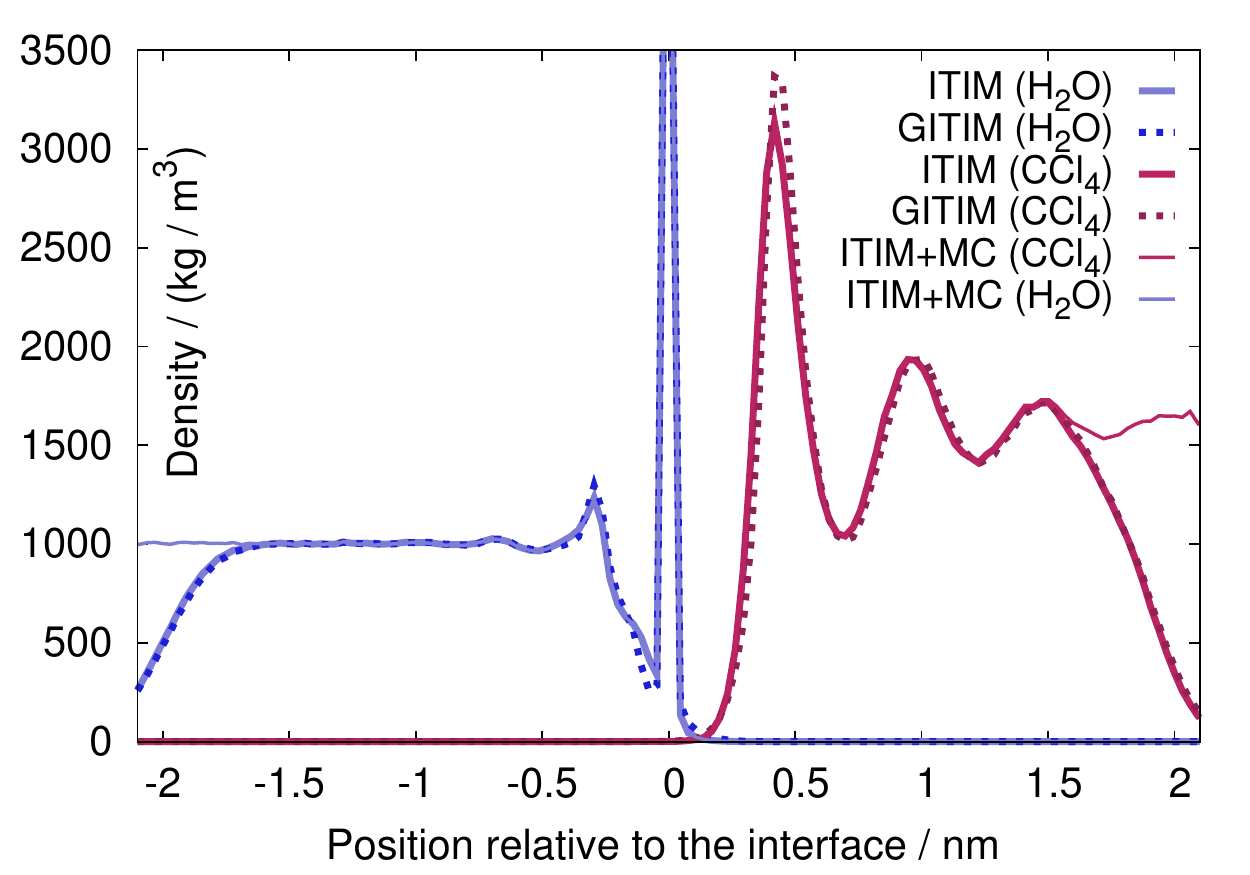}
\caption{
Intrinsic density profiles of water (curves on the left) and carbon tetrachloride (curves on the right) with respect to the water surface as computed with \itim{} (solid curves) or with \gitim{} (dashed curves). The thin curves are computed using \itim{} and the Monte Carlo normalization procedure described in Sec.~\ref{sec:nonplanar}.\label{fig:density}}
\end{figure}

The intrinsic density profiles of water and carbon tetrachloride
are reported in Fig.~\ref{fig:density}, as computed by \itim{}
and \gitim{}, respectively, with the interfacial water molecules
as reference. The procedure for identifying the local distance of
an atom from the surface is in its essence the same as described
in Ref.~\onlinecite{jorge10b}. Starting from the projection $P_0=(x,y)$
of the position of the given atom to the macroscopic interface plane,
the two interfacial atoms closest to $P_0$ are found (their position
on the interface plane being $P_1$ and $P_2$, respectively). The third closest
atom with projection $P_3$ has then to be found, with the condition
that the triangle ${P_1P_2P_3}$ contains the point $P_0$. A linear
interpolation of the elevation of $P_0$ from those of the other
points is eventually performed, and employed to compute the distance
$z-\xi(x,y)$ which is used to compute the intrinsic density profile.

Efficient neighbor search for the $P_1$, $P_2$ and candidate $P_3$
atoms is implemented using kd-trees\cite{bentley79a} as discussed before.
The two pairs of profiles are very similar, besides a small difference
in the position and height of the main peak of the CCl$_4$ profile
(curves on the right in Fig.~\ref{fig:density}) and in the minimum of the water
profile (curves on the left in Fig.~\ref{fig:density}) right next to the
surface position, which are anyway compatible with the differences
observed between various methods for the calculation of intrinsic
density profiles\cite{jorge10b}. The delta-like contribution of the water molecules at the
surface is included in the plot in Fig.~\ref{fig:density}, and defines
the origin of the reference system. Negative values of the
signed distance from the interface correspond to the aqueous phase.

\section{The problem of normalization of density profiles\label{sec:nonplanar}}
Before applying \gitim{} to non-planar interfaces, one important issue
has still to be solved, namely that of the proper calculation of
intrinsic density profiles in non-planar geometries. In general,
one uses one-dimensional density profiles (intrinsic or non-intrinsic)
when the system is, or is assumed to be, invariant under displacements
along the interface, so that the orthogonal degrees of freedom can
be integrated out. When the interface has a non-planar shape, one
needs to use a different coordinate system. In the case of a quasi-spherical object
for example, one could use the spherical
coordinate system to compute the non-intrinsic density profile, and
normalize each bin by the integral of the Jacobian determinant,
that is the volume of the shell at constant distance from the origin.
In the intrinsic case, however, it is necessary to know at every
time step the volume of the shells at constant distance from the
interface. 

The volume of shells at constant intrinsic distance can, in principle,
be calculated at each frame by regular numerical integration,
but this would require a large computing time and storage overhead.
Here, instead, we propose to employ an approach based on simple
Monte Carlo integration: in parallel with the calculation of the
histograms for the various phases, we compute also that of a random
distribution of points, equal in number to the total atoms in the
simulation. The volume of a shell can be estimated as box volume times the ratio of the number of points found at a given distance and the total number of random points drawn. We are following the heuristic idea that for each frame
$j$ one does not need to know the volume of the shell $V_j(r)$ with
a precision higher than that of the average number of atoms in it,
$N_j(r)$. In addition, we assume that the surface area of the
interface is large enough for the shell volume variations $\delta
V_j(r)$ to be small with respect to its average value $\hat{V}(r)=\sum_j^N
V_j(r)/N$. The average density \begin{equation} \rho(r)=
\frac{1}{N}\sum_{j=1}^N \frac{N_j(r)}{V_j(r)} \end{equation} can
be approximated as \begin{equation} \rho(r)\simeq
\frac{1}{N}\frac{1}{\hat{V}(r)} \sum_{j=1}^N \left[ N_j(r) -
N_j(r)\frac{\delta V_j(r)}{\hat{V}(r)} \right].  \end{equation}
When the relative volume changes $|\delta V/V|$ are small, one can
therefore simply normalize the histogram $\hat{N}(r)=\sum_j^N
N_j(r)/N$ by the average volume  $\hat{V}(r)$ obtained by the Monte
Carlo procedure, disregarding the terms of order $\mathcal{O}(\delta V /\hat{V})$.

The correctness of our assumption is demonstrated incidentally by
the application of this normalization once again to the planar case.
The thin lines in Fig.~\ref{fig:density} represent the \itim{}
intrinsic  mass density profile of water and carbon tetrachloride,
using the Monte Carlo normalization scheme instead of the usual
normalization with box cross sectional area and slab width. Close
to the interface, the Monte Carlo normalization gives results which
are fully compatible with the usual method, showing that the accuracy
of the volume estimate is adequate. On the other hand one can see
that far from the interface the two profiles behave quite differently.
The case with usual normalization decays slowly to zero: this effect
is due to the presence of the second interface, whose profile is
smeared again by capillary waves. The case with Monte Carlo
normalization, on the contrary, shows that it is possible to recover
the proper intrinsic density also at larger distances, and features
such as the fourth peak at 2 nm, which are completely hidden in the
normal picture, can be revealed. This shows that the use of the
proper, curvilinear coordinate system is of fundamental importance
also for macroscopically planar interfaces. The calculation of the
Monte Carlo normalization factors does not change the typical scaling
of the algorithm, as it consists in calculating the histogram for
an additional phase of randomly distributed points (which effectively
behaves as an ideal gas).  The better accuracy at larger distances,
however, demonstrates that the use of the Monte Carlo normalization
is much more efficient than the standard approach, as it requires
much smaller systems to be able to extract the same information
(e.g., to resolve the fourth peak in Fig.~\ref{fig:density}, an
additional slab of about 2-3 nm would have been needed). In this
sense, the Monte Carlo normalization procedure can be even beneficial
in terms of performance.

\section{Examples of non-planar interfaces}

\subsection{DPC micelle}

\begin{figure}[t]
\includegraphics[width=0.9\columnwidth]{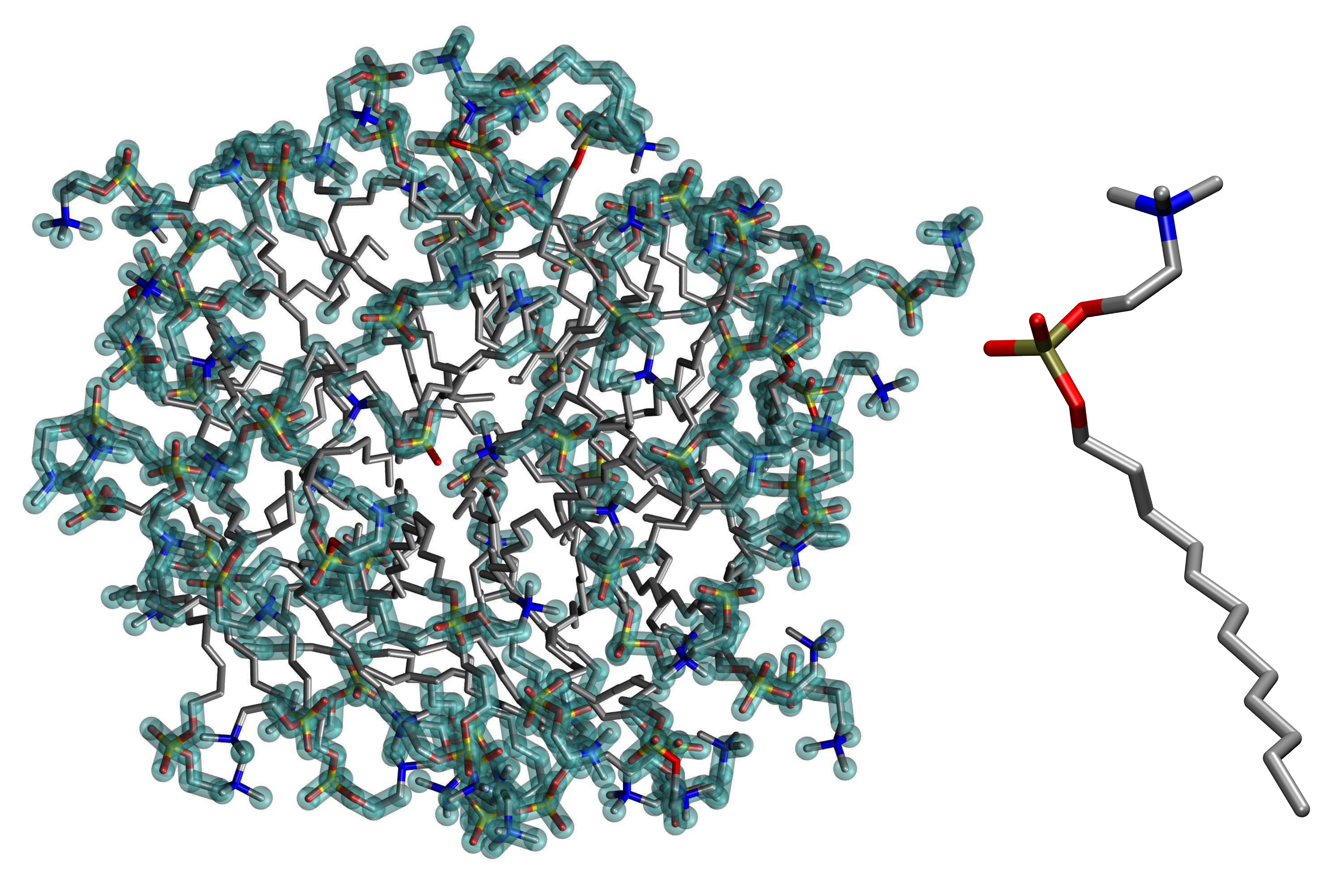}
\caption{Right: schematic structure of a DPC molecule. Left: snapshot of a DPC micelle in water. Only the DPC constituents are shown for the sake of clarity. Atoms with a halo are those recognized by \gitim{} as surface ones.\label{fig:micelle}}
\end{figure}
\begin{figure}[t]
\includegraphics[width=\columnwidth]{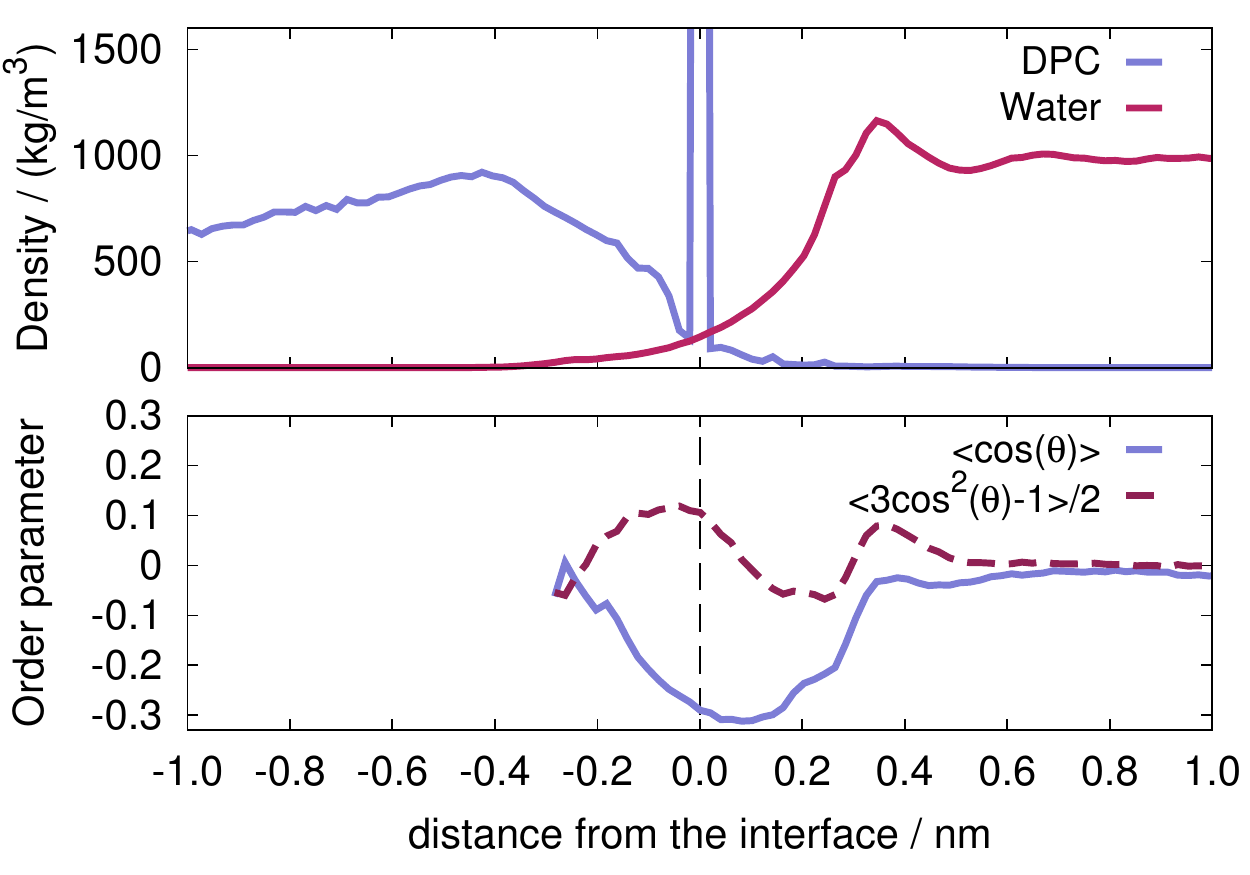}
\caption{Upper panel: intrinsic density profiles of water (right) and of DPC (left). Lower panel: intrinsic profile of the orientational order parameters ($P_1$, solid line, $P_2$, dashed line). The vertical dashed lines marks the position of the interface. \label{fig:order}}
\end{figure}

Dodecylphosphocholine (DPC) is a neutral, amphiphilic molecule with
a single fatty tail that can form micelles in solution: these play
a relevant role in biochemistry, especially for NMR spectroscopy
investigations aiming at understanding the structure of proteins
or peptides bound to  an environment that is similar to the
biological membrane\cite{rozek00a,gesell97a,schibli01a,kallick95a}. The molecular structure of DPC is shown in Fig.\ref{fig:micelle}.
We have simulated for 500 ps a micelle of 65 DPC and 6305
water molecules using the force field and configurations from Tieleman and colleagues\cite{tieleman00a}, and have calculated
the intrinsic mass density profiles of both phases (DPC and water)
using \gitim{} and the Monte Carlo normalization procedure, with a
probe sphere radius $R_p=0.25$. The result of the interfacial atoms
identification on the DPC micelle for a single frame is shown in
Fig.~\ref{fig:micelle}, where water molecules have been removed for
the sake of clarity, and interfacial atoms are highlighted as usual
with a halo. The intrinsic mass density profile, calculated relative to the DPC surface, is reported in
Fig.~\ref{fig:order}, with the DPC mass density profile shown on
the left, and the water profile on the right. 

As usual, the delta-like
contribution at $r=0$ identifies the contribution
from interfacial DPC atoms. In addition, we have calculated, for the first time, the intrinsic profiles of the orientational order
parameters $S_1$ and $S_2$ of the water molecules around the DPC
micelle. The two parameters are defined as $S_1=\left\langle
\cos(\theta_1)  \right\rangle$ and $S_2=\left\langle  3\cos^2(\theta_2)
-1 \right\rangle/2$, where $\theta_1$ and $\theta_2$ are the angles
between the water molecule position vector (with respect to the
micelle center), and the water molecule symmetry axis and molecular
plane normal, respectively. The orientation is taken so that
$\theta_1<\pi/2$ when the hydrogen atoms are farther from the micelle
than the corresponding oxygen. The complete picture of the orientation
of water molecules would be delivered by the calculation of the
probability distribution $p(\theta_1,\theta_2)$\cite{jedlovszky02a,jedlovszky04a}, but here we limit
our analysis to the two separate order parameters and their intrinsic
profiles. Note that, since these quantities are computed per particle,
there is no need to apply any volume normalization. The polarization
of water molecules, which is proportional to $S_1$, appears to be
different from zero  only very close to the micellar surface. In
particular, $S_1$ has a correlation with the main peak of the
intrinsic density profile in the proximity of 0.4 nm. Water molecules
located closer to the interface show a first change in the sign of
the polarization and a subsequent one when crossing the interface.
Farther than 0.25 nm inside the micelle, not enough water molecules are found to generate any
meaningful statistics. Also the $S_2$ order parameter is practically
zero beyond 0.6 nm, and again a correlation is seen with the main
peak of the intrinsic density profile, and the maximum in the
orientational preference is found just next to the interface,
where $S_1\simeq 0$, showing that water molecules are preferentially
laying parallel to the interfacial surface.

\subsection{Soot}

\begin{figure}[t!]
\includegraphics[width=\columnwidth]{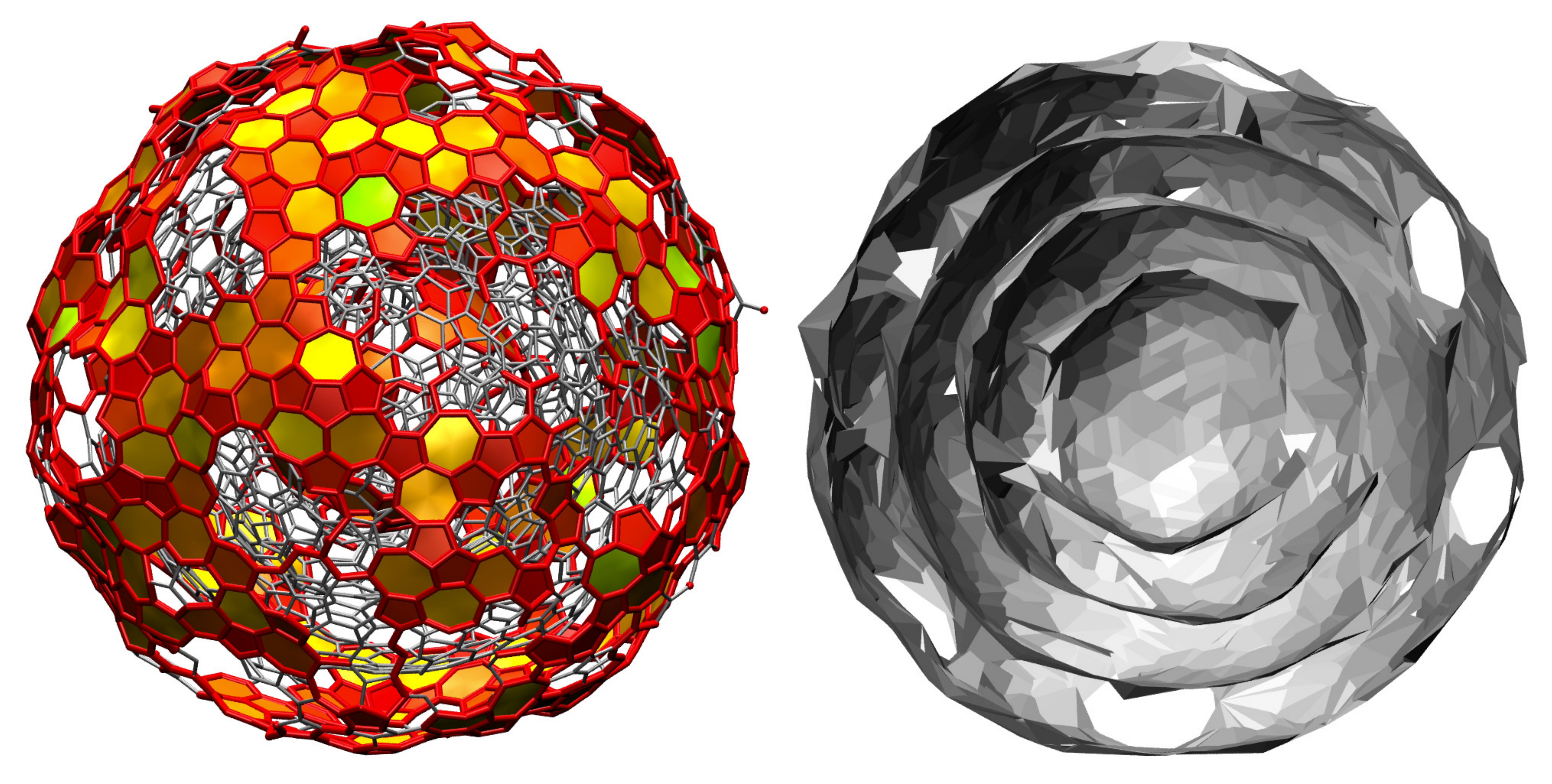}
\caption{The S$^\mathrm{I}_\mathrm{1}$ soot model\cite{hantal10a} represented in section (right, triangulated surface) and in whole (left, wireframe) with the atoms identified by \gitim{} as surface ones highlighted using thicker, red elements. Besides surface atoms, also chemical bonds between surface atoms are highlighted, as well as five, six and seven membered rings (filled surfaces).\label{fig:soot}}
\end{figure}

\begin{figure}[t!]
\includegraphics[width=\columnwidth]{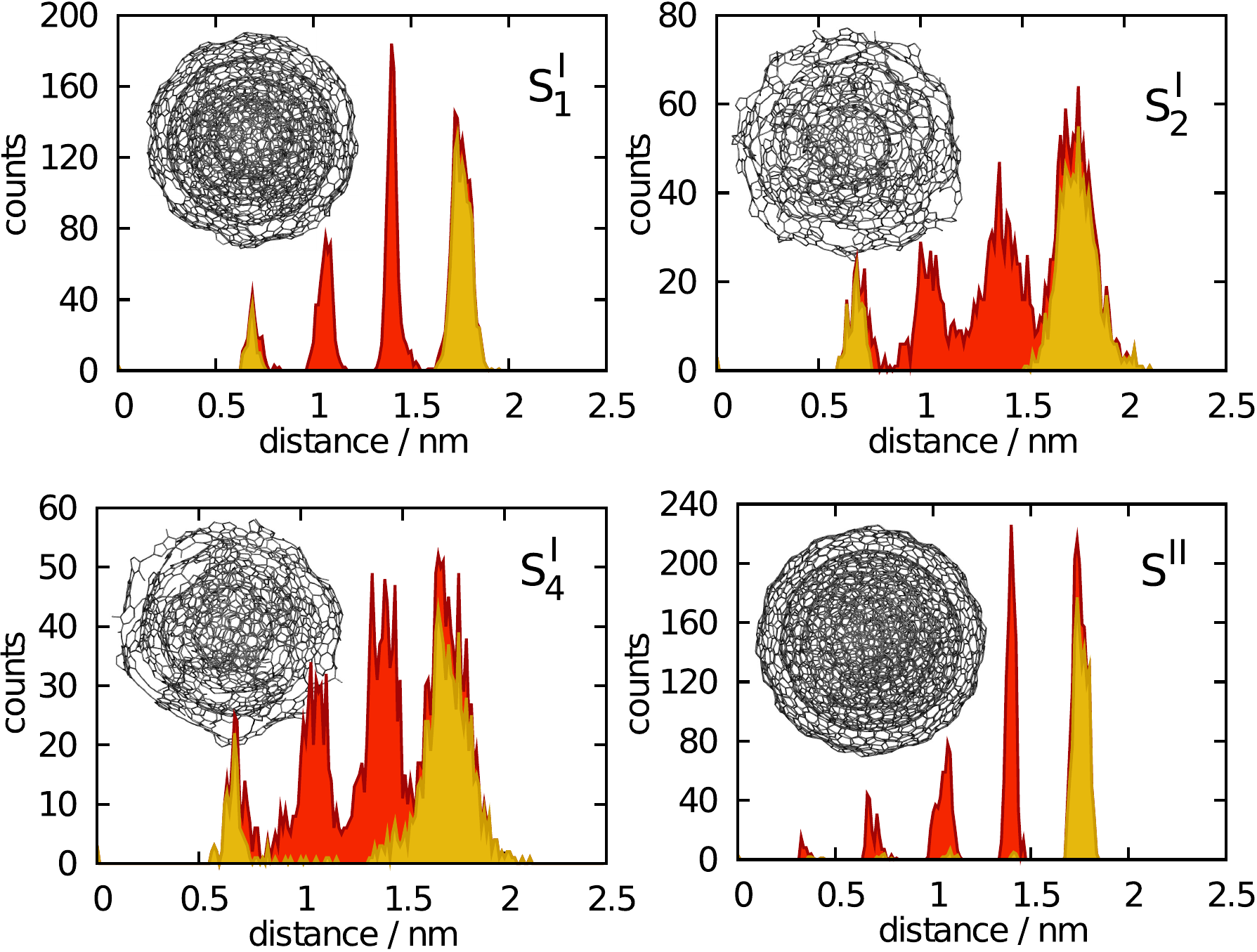}
\caption{Histograms of the atoms in the four soot models taken from Ref.~\onlinecite{hantal10a}. Each panel refers to a different structure (depicted with wireframe), and presents the distribution of all atoms (filled, darker area) and of surface atoms identified by \gitim{} (filled, lighter area), as a function of the distance from the center.\label{fig:sootdens}}
\end{figure}

One of the main byproducts of hydrocarbon flames, soot is thought
to have a relevant impact on atmospheric chemistry and global surface
warming\cite{quaas11a,vanrenssen12a}. Electron, UV, and atomic force
microscopy have revealed the size and structure of soot particles from
different sources at different
scales\cite{popovitcheva00a,sgro03a,chen05a,abid08a,abid09a}. 
In particular, soot emitted by aircraft is found to be made of
several, quasi-spherical, concentric
graphitic layers of size in the range from 5 to 50
nm\cite{popovitcheva00a}.  We have used four model structures
(S$^\mathrm{I}_\mathrm{1}$,S$^\mathrm{I}_\mathrm{2}$,
S$^\mathrm{I}_\mathrm{4}$ and S$^\mathrm{II}$ from
Ref.~\onlinecite{hantal10a}) to demonstrate the ability of \gitim{}
to identify surface atoms in complex geometries. In Fig.~\ref{fig:soot},
the S$^\mathrm{I}_\mathrm{1}$ model is represented in section as a
triangulated surface (right), showing the four concentric layers,
and in whole (left) showing the surface atoms as detected by \gitim{}
using $R_p=0.25$ nm. The histograms of the total number of atoms
and of the surface ones, as a function of the distance from the
center of the soot particles, are shown in Fig.~\ref{fig:sootdens}
for the four different models, where it is seen how particles of
the size of a water molecule have mostly access only to the inner and outer
parts of the innermost and outermost shell, respectively, and cover
them almost completely. This finding is in a clear accordance with
the results of the void analysis and adsorption isotherm calculations
presented in Ref.~\onlinecite{hantal10a}.

\subsection{Secondary cholic acid micelle}

Bile acids, such as cholic acid are biological amphiphiles built
up by a steroid skeleton and side groups attached to it. The
organization of these side groups is such that hydrophilic and
hydrophobic groups are located at the two opposite sides of the
steroid ring. Thus, bile acids have a hydrophilic and a hydrophobic
face (often referred to as the $\alpha$ and $\beta$ side, respectively)
rather than a polar head and an apolar tail, as in the case of other surfactants like, for example, DPC. The unusual molecular shape leads to peculiar
aggregation behavior of bile acids. At relatively low concentrations
they form regular micelles with an aggregation number of 2-10,
while above a second critical micellar concentration these primary
micelles form larger secondary aggregates by establishing hydrogen
bonds between the hydrophilic surface groups of the primary micelles
\cite{small71a,partay07a}. These secondary micelles are of rather
irregular shape,\cite{partay07a,partay07b} which makes them an
excellent test system for our purposes.

\begin{figure}[t!]
\includegraphics[width=\columnwidth]{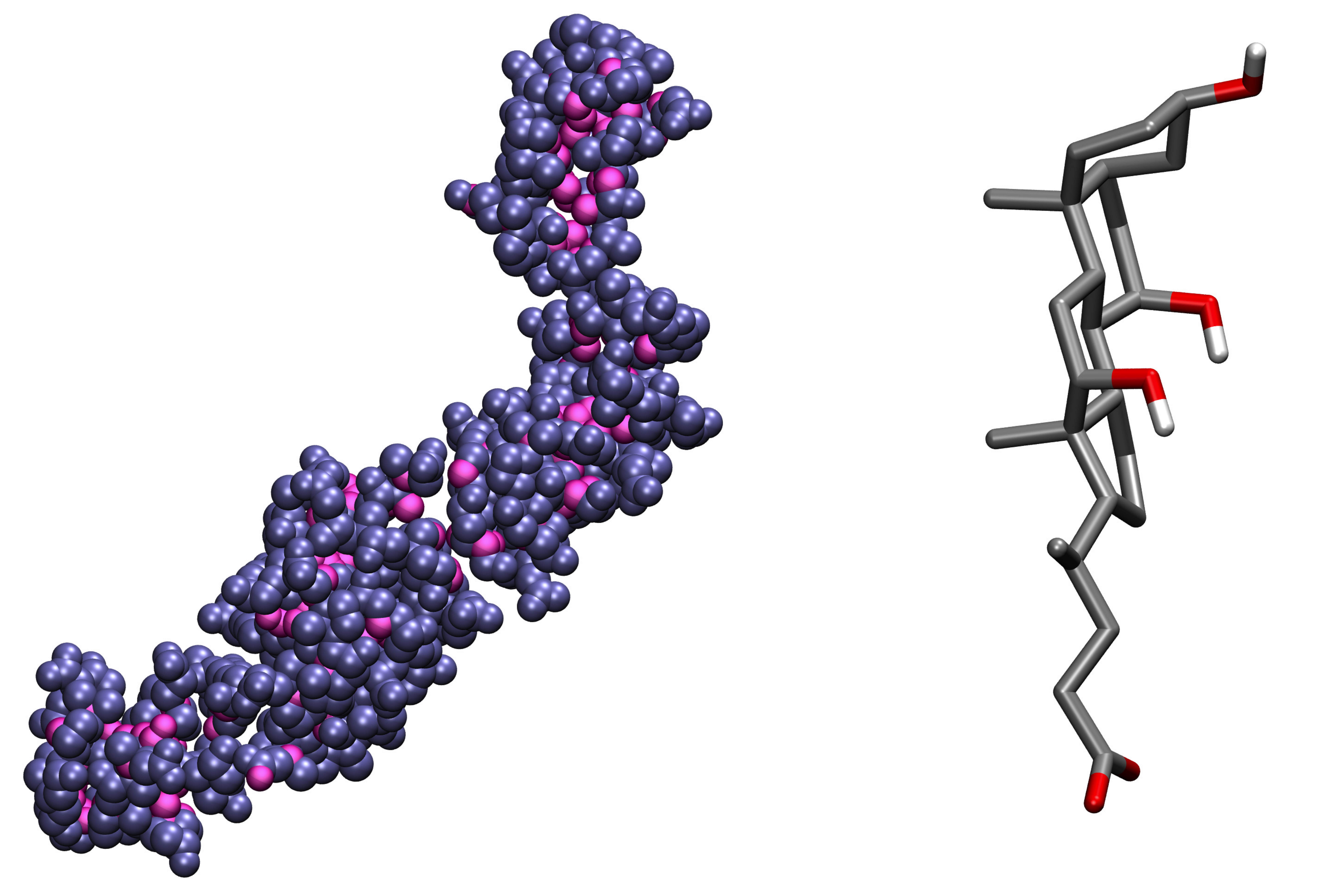}
\caption{Left: simulation snapshot of a secondary cholate micelle, with surface atoms highlighted. Right: the structure of the cholic acid molecule.\label{fig:bile}}
\end{figure}

\begin{figure}[t!]
\includegraphics[width=\columnwidth]{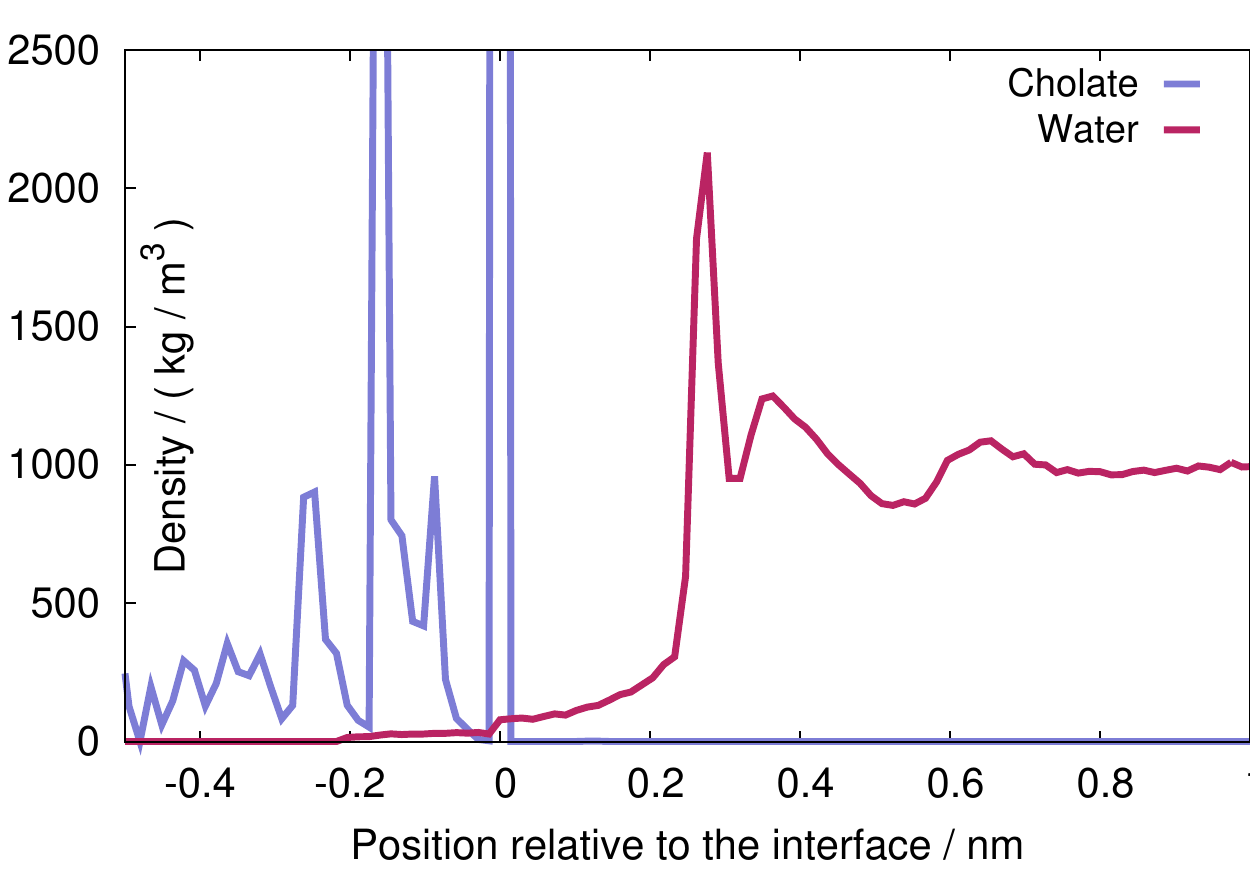}
\caption{Density profile of water (right) and cholic acid (left) in the secondary micelle.\label{fig:biledens}}
\end{figure}

Here we analyze the surface of a secondary cholic acid micelle
composed of 35 molecules, extracted from a previous simulation
work\cite{partay07a} and simulated for the present purposes for 500
ps in aqueous environment. An instantaneous snapshot of the micelle is shown in
Fig.\ref{fig:bile} (water molecules are omitted for clarity) together
with a schematic structure of the cholic acid molecule. We calculated
the density profile of water as well as of cholic acid relative to
the intrinsic surface of the micelle by the \gitim{} method. The
resulting profiles are shown in Fig.\ref{fig:biledens}.
The micelle has a characteristic elongated shape, which exposes a
large part of its components to the solvent, so that roughly 80\%
of the micelle atoms are identified as surface ones. The small
volume to surface ratio of the micelle is at the origin of the
rather noisy intrinsic density profile for the micelle itself. The
profile, in addition to the delta-like contribution at the surface,
presents another very sharp peak located at a distance of about
0.18 nm inside the surface, due to the rather rigid structure of
the bile molecule. The water intrinsic density profile, on the
contrary, shows a marked peak at 0.25 nm, absent in the DPC micelle
case, due to the presence of hydrogen bonds between water molecules
and the hydroxyl groups of cholic acid.


\section{Conclusions}

In this paper we presented a new algorithm that combines the
advantageous features of both the \itim{} method \cite{partay08a}
and the $\alpha$-shapes algorithm\cite{edelsbrunner83a,edelsbrunner94a}
to be used in determining the intrinsic surface in molecular
simulations. Thus, unlike the original variant, this new, generalized
version of \itim{}, dubbed \gitim{}, is able to treat interfaces
of arbitrary shapes and, at the same time, to take into account the
excluded volumes of the atoms in the system.  It should be emphasized
that the \gitim{} algorithm is not only able to find the external
surface of the phase of interest, but it also detects the surface
of possible internal voids inside the phase.  The method, based on
inflating probe spheres up to a certain radius in points inside the
phase turned out to provide practically identical results with the
original \itim{} analysis for planar interfaces.  Further, its
applicability to non-planar interfaces was shown for three previously
simulated systems, i.e., a quasi-spherical micelle of
DPC\cite{tieleman00a}, molecular models of soot\cite{hantal10a},
and a secondary micellar aggregate of irregular shape built up by
cholic acid molecules \cite{partay07a}.

Another important result of this paper concerns the correct way of
calculating density profiles relative to intrinsic interfaces,
irrespective of whether they are macroscopically planar or not.
Thus, here we proposed a Monte Carlo-based integration algorithm
to estimate the volume elements in which points of the profile are
calculated, in order to normalize them correctly. The issue of
normalization with the volume elements in macroscopically flat fluid
interfaces originates from the fact that these interfaces are rough
on the molecular scale, namely, at the length scale of the calculated
profiles. We clearly demonstrated that using this new 
normalization the artificial smearing of the intrinsic density
profiles far from the intrinsic interface can be avoided.

Two computer programs that implement, respectively, an optimized
version of \itim{} and the new \gitim{} algorithm, as well as the
calculation of intrinsic density and order parameters profiles, are
made available free of charge at \verb!http://www.gitim.eu/!.
The programs are compatible with the trajectory and topology file
formats of the Gromacs molecular simulation package~\cite{hess08a}.

\section*{Acknowledgements}
M.S. acknowledges FP7-IDEAS-ERC grant DROEMU: ``Droplets \& Emulsions: Dynamics
\& Rheology'' for financial support, and Gy\"orgy Hantal for providing
the soot structures. Part of this work has been done by M.S. at
ICP, Stuttgart University. P.J. is grateful for financial support
from the Hungarian OTKA foundation under Project Nr. 104234.  S.K.
is supported by FP7-IDEAS-ERC Grant PATCHYCOLLOIDS 
and RFBR Grant mol\_a 12-02-31374. Simulation snapshots were made
with VMD\cite{humphrey96a}.

\section{Appendix}

Here, following  Ref.~\onlinecite{penfold05}  we derive the expressions
for the radius $R$ and position $\mathbf{r}=(x,y,z)$ of the center
of the sphere which is touching four other ones, having given radii
and center positions  $R_i$ and $\mathbf{r}_i=(x_i,y_i,z_i)$
($i=1,2,3$ or 4), respectively.  The conditions of touching can be
expressed with the following nonlinear system of four equations:

\begin{equation}
|\mathbf{r}-\mathbf{r}_i|^2=(R+R_i)^2.\label{eq:constr}
\end{equation}
By subtracting  one of them from the other three (without loss of generality we subtract the one with $i=1$), the quadratic term, $\mathbf{r}^2$, will be eliminated and the system Eq.(\ref{eq:constr})
would become linear with respect to $\mathbf{r}$:
\begin{equation}
\mathbf{M}\mathbf{r} = \mathbf{s} - R \mathbf{d},\label{eq:linear}
\end{equation}
where the matrix $\mathbf{M}$ and the vectors $\mathbf{d}$ and $\mathbf{s}$ are defined as
\begin{eqnarray}
\mathbf{M}=\left( 
\begin{array}{c}
\mathbf{r}_1-\mathbf{r}_2\\
\mathbf{r}_1-\mathbf{r}_3\\
\mathbf{r}_1-\mathbf{r}_4\\
\end{array}
\right),
& 
\mathbf{d}=\left( 
\begin{array}{c}
R_1-R_2\\
R_1-R_3\\
R_1-R_4\\
\end{array}
\right),
\end{eqnarray}
and
\begin{equation}
\mathbf{s}=\frac{1}{2}
\left(
\begin{array}{c}
\mathbf{r}_1^2-\mathbf{r}_2^2-R_1^2+R_2^2 \\[0.2em] 
\mathbf{r}_1^2-\mathbf{r}_3^2-R_1^2+R_3^2 \\[0.2em]
\mathbf{r}_1^2-\mathbf{r}_4^2-R_1^2+R_4^2 \\[0.2em]
\end{array}
\right).
\end{equation}
Equation (\ref{eq:linear}) has a unique solution if matrix $\mathbf{M}$ is non-singular (the singularity of $\mathbf{M}$ corresponds to the case when all 4 spheres are co-planar, which means that the unknown sphere either does not exist, or is not unique):
\begin{equation}
\mathbf{r}=\mathbf{M}^{-1}\mathbf{s}-R\mathbf{M}^{-1}\mathbf{d}\equiv \mathbf{r}_0-R\mathbf{u},\label{eq:r}
\end{equation}
where $\mathbf{M}^{-1}\mathbf{s}=\mathbf{r}_0$ and $\mathbf{u}=\mathbf{M}^{-1}\mathbf{d}$.  Once Eq.(\ref{eq:r}) is substituted into the first of the constraints Eq.(\ref{eq:constr}), it leads to the quadratic algebraic equation with respect to $R$: 
\begin{equation}
\left(1-|\mathbf{u}|^2\right)R^2 + 2 \left(R_1-\mathbf{u}\cdot\mathbf{v}\right)R + (R_1^2-|\mathbf{v}|^2) = 0,\label{eq:2ndorder}
\end{equation}
where $\mathbf{v}=\mathbf{r}_1-\mathbf{r}_0$. The solution of Eq. (\ref{eq:2ndorder}) can be found in the following form:
\begin{equation}
\label{eq:R_pm}
R_{\pm} =  \frac{-\left(R_1-\mathbf{u}\cdot\mathbf{v}\right)\pm |R_1\mathbf{u} +\mathbf{v}|}{1-|\mathbf{u}|^2}.
\end{equation}

\noindent If $|\mathbf{u}|^2$ is not equal to unity (which corresponds
to the case when the 4 spheres are tangential to one plane), then
Eq.(\ref{eq:R_pm}) provides two different solutions, and the positive
one provides the radius $R$ of the touching sphere as a function
of the centre position $\mathbf{r}$. Eventually, the positions of
their centres can be obtained by inserting $R$ into Eq.(\ref{eq:r}).
In the present manuscript in case of two possible solutions we
choose the sphere with minimal radius.

\makeatletter
\renewcommand{\@dotsep}{2.5}
\makeatother
\clearpage

\clearpage
\end{document}